# GRAVITATIONAL COLLAPSE OF SMALL-SCALE STRUCTURE AS THE ORIGIN OF THE LYMAN ALPHA FOREST

Renyue Cen[1], Jordi Miralda-Escudé[2], Jeremiah P. Ostriker[1], and Michael Rauch[3]

email: cen@astro.princeton.edu

[1] Princeton University Observatory, Princeton, NJ 08544

[2] Institute for Advanced Study, Princeton, NJ 08540

[3] Observatories of the Carnegie Institution of Washington, Pasadena, CA 91101

ABSTRACT

If gravitational clustering is a hierarchical process, the present large-scale structure of the galaxy distribution implies that structures on smaller scales must have formed at high redshift. We simulate the formation of small-scale structure (average cell mass: $\Delta \bar{m}_b = 10^{4.2} M_\odot$) and the evolution of photoionized gas, in the specific case of a CDM model with a cosmological constant. The photoionized gas has a natural minimal scale of collapse, the Jeans scale ($m_{b,J} \simeq 10^9 M_\odot$). We find that low column density ($N_{HI} \leq 10^{14} \mathrm{cm}^{-2}$) lines originate in regions resembling Zel'dovich pancakes, where gas with overdensities in the range $3 - 30$ is enclosed by two shocks but is typically re-expanding at approximately the Hubble velocity. However, higher column density ($N_{HI} \geq 10^{15} \mathrm{cm}^{-2}$) lines stem from more overdense regions where the shocked gas is cooling. We show that this model can probably account for the observed number of lines, their distribution in column density and b-parameters, as well as the cloud physical sizes as observed in gravitationally lensed quasars. We find a redshift evolution that is too steep; however, this may be due to insufficient dynamical range in the simulation or because the specific model (CDM+$\Lambda$) is incorrect. The model predicts that high signal-to-noise observations should find systematic deviations from Voigt profiles, mainly in the form of broad wings in the line profiles, and that a fluctuating Gunn-Peterson effect will be detected, which can be modeled as a superposition of weak lines with a wide range of b-parameters.

Cosmology: early universe –Cosmology: large-scale structure of Universe – hydrodynamics – quasars: absorption lines

# 1. INTRODUCTION

The physical origin of the $Ly_\alpha$ forest in the QSO absorption spectra has remained a mystery. The essential, difficult fact is that the clouds have "size" $\sim 10 - 30$kpc and are typically separated by $\sim 2,000$kpc along the line of sight, but produce most of the absorption. Thus, the absorption per unit length within a cloud must exceed the absorption per unit length in the intercloud medium by a factor of more than $10^2$. Because the neutral hydrogen is proportional to $\rho^2$, the clouds must then be overdense by a factor $> 10$. This has led to extensive discussions of the possible confinement mechanism, with gravity and external gas pressure being proposed, as well as free expansion. We follow up the scenario of gravitational collapse on subgalactic scales for the origin of the $Ly_\alpha$ clouds (Ikeuchi 1986; Rees 1986; Bond, Szalay, & Silk 1988). We go beyond the previous qualitative models, by performing direct, high resolution, numerical hydro simulations of gravitational collapse on small scales. We find that we do, in fact, reproduce most of the observed properties of the $Ly_\alpha$ forest in a simulation of the CDM+$\Lambda$ model, although, as we shall see, the physical picture is different from that anticipated in prior semi-analytic calculations.

If this theory of the $Ly_\alpha$ forest is correct, then the observations of $Ly_\alpha$ extinction in quasars will provide a new, direct probe to the initial density fluctuations in the universe. Such observations have three advantages when compared to local ($z = 0$) observations of galaxies for the assessment of cosmological models: (1) they are unbiased in that they probe random lines of sight; (2) they can reach to relatively high redshift ($z \geq 4$), where the contrasts between competing models are greatest; (3) they provide a large database with $\sim 10^2$ lines/quasar $\times 10^3$ quasars $\sim 10^5$ lines as the potential data base.

# 2. NUMERICAL MODELING

We use the new shock-capturing cosmological hydrodynamic code described in Ryu et al. (1993) called Total Variation Diminishing (TVD) with addition of atomic processes within a primeval plasma of (H,He). We calculate self-consistently the average background photoionizing radiation field as a function of frequency. Ionization equilibrium of the abundances of HI, HII, HeI, HeII, and HeIII is required at every timestep, but the time evolution of the temperature in every cell is solved explicitly, using the same heating and cooling terms as in Cen (1992).

We model galaxy formation as in Cen & Ostriker (1992, 1993a,b). The material turning into galaxies is assumed to emit ionizing radiation, with two types of spectra: one characteristic of star formation regions and the other characteristic of quasars, with efficiencies (i.e., the fraction of rest-mass energy converted into radiation) of $e_{UV,*} = 5 \times 10^{-6}$, and $e_{UV,Q} = 6 \times 10^{-6}$, respectively. The derived intensity of the radiation field at the Lyman limit is $J = (2.3, 3.4, 1.8) \times 10^{-21} \text{erg cm}^{-2}\text{sec}^{-1}\text{hz}^{-1}\text{sr}^{-1}$ at $z = (2, 3, 4)$, slightly higher than suggested by observations of the proximity effect (Murdoch et al. 1986; Bajtlik, Duncan, & Ostriker 1988; Lu et al. 1991).

We adopt a CDM+$\Lambda$ model with the following parameters: $\Omega = 0.4$, $\Omega_b = 0.0355$ (cf. Walker et al. 1991), $\lambda = 0.6$, $h = 0.65$ and $\sigma_8 = 0.79$ (COBE normalization; Kofman, Gnedin, & Bahcall 1993). This model is consistent with all observational tests that we know of. Nevertheless, we have no reason to believe that it is, in fact, correct. Our box size is $3h^{-1}$Mpc with $N = 288^3$ cells and $144^3$ dark matter particles. The cell size is $10.4h^{-1}$kpc (comoving) corresponding to a baryonic mass of $1.7 \times 10^4 M_\odot$, with true spatial resolution slightly worse than this. At redshift three, the Jeans length, $\lambda_J \equiv (\pi c_s^2/G\bar{\rho}_{tot})^{1/2}$ for $c_s = v_{rms} = 10\text{kms}^{-1}$, is equal to $400h^{-1}$kpc in comoving units, or 40 cells, so the smallest scales of initial collapse for the photoionized gas are well resolved.

3. PHYSICAL PROPERTIES OF THE COLLAPSED STRUCTURES

Figure 1 shows contour plots of four quantities projected accross a slice of the simulation of thickness $0.5h^{-1}$Mpc: total gas density ($\rho_b$), neutral hydrogen column density accross the slice ($\rho_{HI}$), dark matter density ($\rho_d$), and gas temperature ($T$). We see that high column density $Ly_\alpha$ clouds ($N_{HI} > 10^{14}\text{cm}^{-2}$) are typically isolated regions, while lower column density clouds are generally in connected filaments and sheets. As expected, photoionization equilibrium prevails so that $T \sim 10^4$K and $n_{HI} \propto n_{Htot}^2$.

Next, we take a random row along the simulation, and examine the physical structures that are intercepted. The upper panel in Figure 2 shows $T$ (solid line) and $\rho_b$ (thick dotted line) along such a row, as a function of the spatial coordinate, given as $v = xH(H = 512\,h\,\mathrm{kms}^{-1}\mathrm{Mpc}^{-1}$ at $z = 3)$. The thin dotted line is the gas pressure ($p$), in arbitrary units. The highest density peak seen resembles the structure of a Zel'dovich pancake (Zel'dovich 1970; Sunyaev & Zel'dovich 1972), with two shocks propagating outwards, and a minimum of the temperature at the center, where the gas has cooled after being shocked. The central minimum exists even in an adiabatic calculation as the shock velocity (and thus the post shock temperature) grows with time.

Figure 1. Coutour plots of four quantities in a slice of $1.5 \times 1.5 \times 0.5 h^{-3} \mathrm{Mpc}^3$ (comoving) at redshift three: density, neutral hydrogen column density, dark matter density and baryonic temperature. Contour levels are $10^{0,1,2}$ for the total gas density and dark matter density, each in terms of its own global mean, and $10^{13+i} \mathrm{cm}^{-2}$ for the neutral hydrogen column density ($i = 0, 1, 2, ...$). The contour levels for temperature are $10^{4+0.1i}$.

We can now re-examine the question of the dynamical mechanism that causes the gas overdensity, and therefore the absorption lines. The possible mechanisms are: (i) Pressure confinement, as proposed by Sargent et al. (1980) and Ostriker & Ikeuchi (1983) does not play a role, since the temperature is higher in the density peaks than in the surrounding medium. (ii) The inertia of the gas; the gas must then be in free expansion (Ikeuchi & Ostriker 1986; Bond, Szalay, & Silk 1988). (iii) Gravity (Rees 1986; Ikeuchi 1986; Ikeuchi & Ostriker 1986). (iv) Ram pressure by infalling gas, which has not been proposed so far. The latter two processes seem to be dominant.

To see this we plot in Figure 2b the gravitational acceleration divided by the Hubble constant, $-\frac{1}{H}\frac{d\phi}{dx}$, as the dashed line, and the total acceleration, $-(\frac{1}{H}\frac{d\phi}{dx} + \frac{1}{\rho}\frac{dp}{dx})$, as the solid line. The dotted line is the peculiar velocity ($v_p$) along the row; in the linear regime, and for $\Omega = 1$, this is given by $v_p = -\frac{2}{3}\frac{d\phi}{dx}$. In the highest density peak in the figure, the spikes in the total acceleration and the discontinuities in the velocity correspond to the two shock fronts. Between the shocks, there is a large gravitational acceleration, and the solid line shows that this is approximately balanced by the pressure gradient. The ram pressure of the infalling gas at the shock is, in this case, about one third of the maximum pressure in the system, so the ram pressure and gravitational binding forces are comparable. We have examined a total of 30 random rows and find that this situation is typical (although the balance between gravitational and pressure forces between shocks is not generally perfect), and the gas is typically expanding in comoving coordinates. In other rows, where the overdensity is higher than 300, in at least one cell (and $N_{HI} \geq 10^{15} \text{cm}^{-2}$), the gas is typically contracting and cooling due to collisional excitation processes.

We then conclude that, although the $Ly_\alpha$ clouds that we simulate arise from gravita-tional collapse, as in Bond, Silk & Szalay (1988) and the minihalo model of Rees (1986) and Ikeuchi (1986), they differ from this model in three important respects: first, most of the absorption lines do not correspond to virialized regions with the gas in hydrostatic equilibrium, but rather arise from the infalling gas forming the small-scale structure (Miralda-Escudé & Rees 1993). Second, the gas of low column density clouds ($N_{HI} \leq 10^{14} \text{cm}^{-2}$) has been shocked, and the ram pressure at the shocks plays an important role in maintaining the overdense region that causes the absorption line, with the post shock gas expanding as in a Zel'dovich pancake. Third, the gas does not need to be in perfect thermal equilibrium with the ionizing background, because it is shock-heated and, depending on the intensity and spectrum of the ionizing background, the cooling time at the typical densities of the clouds can be comparable to the Hubble time (Efstathiou 1992). Shocks will be quasi-isothermal with the consequence that large density jump $(\rho'/\rho) \approx (\gamma+1)/(\gamma-1)$ occurs in the transition.

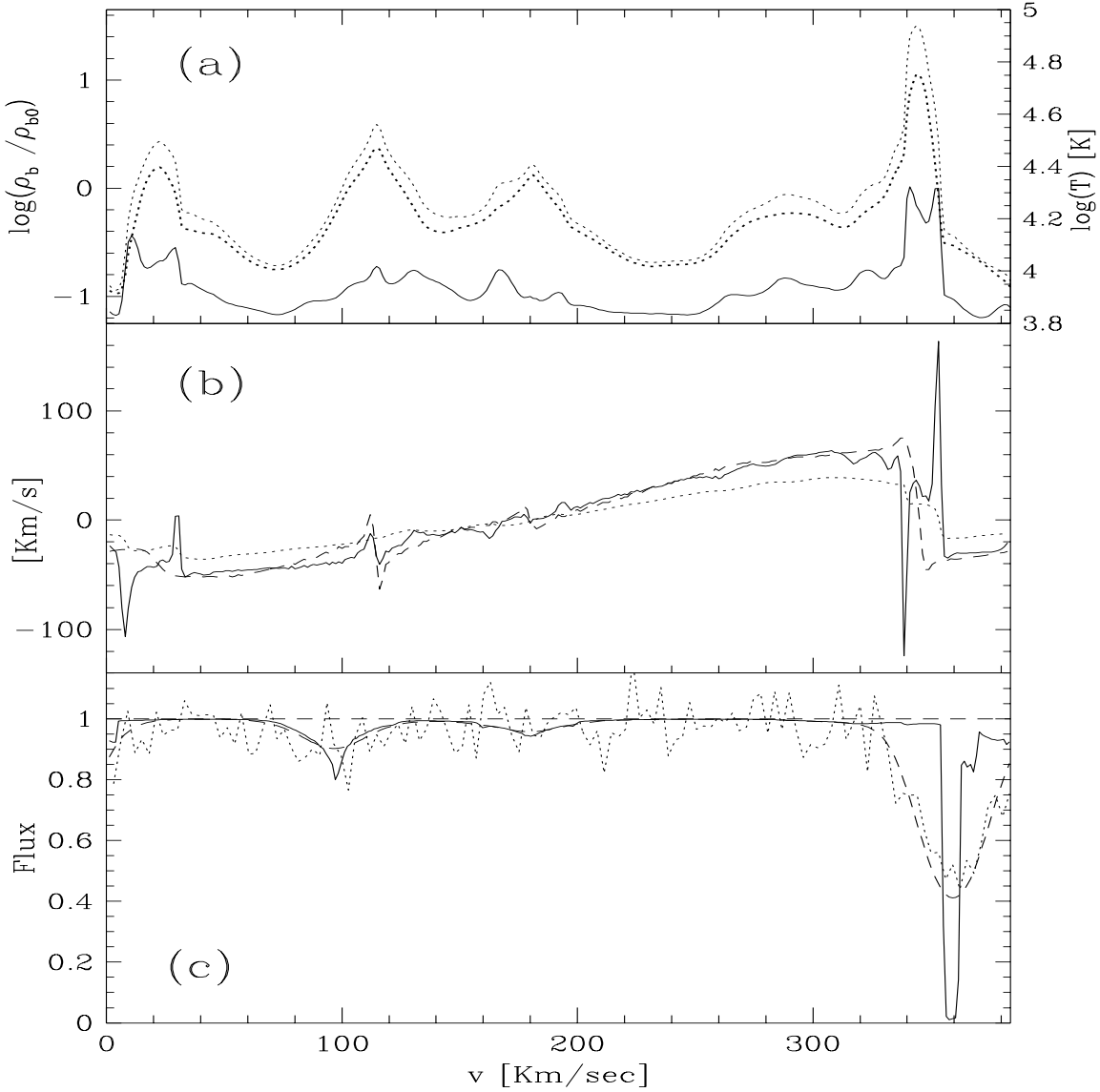

Figure 2. Panel (a) shows the gas temperature (solid line) and gas density (thick dotted line) along such a row, as a function of the spatial coordinate, given as $v = x\,H$. The thin dotted line is $p$, in arbitrary units. Panel (b) shows the gravitational acceleration divided by the Hubble constant, $-\frac{1}{H}\frac{d\phi}{dx}$, as the dashed line, and the total acceleration $-\frac{1}{H}(\frac{d\phi}{dx} + \frac{1}{\rho}\frac{dp}{dx})$, as the solid line. The dotted line is the peculiar velocity along the row. Panel (c) shows the flux reduction from unity for the row featured in Figure 2a plotted as the solid line. The spectrum convolved with a gaussian of width $b(x) \equiv [2kT(x)/m_H]^{1/2}$ is shown as the dashed curve. The dotted curve is obtained by convolving the dashed curve with a gaussian of $FWHM = 10$ kms$^{-1}$ to simulate the instrumental resolution, with added gaussian noise having $S/N = 15$ on a sampling bin of $3$ kms$^{-1}$ (e.g., Carswell et al. 1991).

We now use $(\rho_{HI}, v_p)$ to calculate the optical depth due to $Ly_\alpha$ scattering, if there were no thermal motions of the hydrogen atoms: $\tau(V) = \tau_{GP} \sum_i \{1 + \delta_{HI}[x_i(V)]\} \frac{dx_i}{dV}$. Here, $V$ is the component of the total velocity of the gas along the line of sight, with respect to an arbitrary origin; $x_i(V)$ are the set of all the positions along the line of sight where the gas moves with velocity $V$; and $\delta_{HI}$ is the overdensity of neutral hydrogen. In the linear regime, there is only one spatial position $x_1$ for every velocity $V$, but when structures start collapsing velocity caustics appear, and within those $x(V)$ is multivalued. The normalization $\tau_{GP}$ (Gunn-Peterson optical depth) is the optical depth if the medium had constant density. Thus, $\tau_{GP} = 4.6 \times 10^5 \, \Omega_b \, h \, y \, (1+z)^3 / H(z)$, where $\Omega_b$ is the baryon density divided by the critical density, $y$ is the neutral fraction of hydrogen, and $H(z)$ is the Hubble constant at redshift $z$. For the present simulation, $\tau_{GP} = 0.052$ at $z = 3$.

The flux reduction from unity ($e^{-\tau}$) for the row featured in Figure 2a is plotted as the solid line in Figure 2c. The resultant spectrum of convolving the optical depth with a gaussian of width $b_{calc}(x) \equiv [2kT(x)/m_H]^{1/2}$ is plotted as a dashed curve in Figure 2c. In general, the absorption lines arise from density peaks, producing narrow and well-defined features. Velocity caustics (see McGill 1990) are not important in generating lines due to thermal broadening.

We now convolve the dashed curve with a gaussian of FWHM= 10 kms$^{-1}$ to simulate the instrumental resolution, and add gaussian noise with $S/N = 15$ (adequate for current observations, e.g., Carswell et al. 1991) on a sampling bin of 3 kms$^{-1}$, obtaining the dotted curve in Figure 2c. We repeat the same procedure on 30 randomly selected rows, and detect absorption features and fit Voigt profiles (cf. Webb 1987 and Carswell et al. 1987) to the simulated spectra in the same way as it is usually done for real data (e.g., Rauch et al. 1992). The statistical significance of any spectral features is determined using the procedure described by Young et al. (1979), with lines being considered real detections when their equivalent width has a probability less than $10^{-5}$ to arise by random fluctuation. Whenever $\chi^2$ has less than 1% probability of occurring by chance, additional components are included to fit the features as blends.

The b-parameters of the lines are in the range $15 - 30$kms$^{-1}$ ($\bar{b}_{calc} = 20$kms$^{-1}$), slightly lower than observed, the later being in the range $15 - 60$ kms$^{-1}$ ($\bar{b}_{obs} \simeq 35$kms$^{-1}$) (cf. Press & Rybicki 1993, Rauch et al. 1992 for values and references to earlier literature). There are two effects that could increase the b-parameters: first, the temperature of the intergalactic medium (8000K° in this simulation) could be higher depending on the epoch of reionization and the spectrum of the ionizing background (Miralda-Escudé & Rees 1994). Second, our small box simulation does not include the large-scale power, which would cause broader lines. In addition, there is evidence that the b-parameters are lower at high redshift (Williger et al. 1994), in better agreement with our derived values. In any case, bulk motions and shock-heating are inherent to our model, and imply that b-parameters should be larger than for photoionized gas in thermal equilibrium, as is observed.

Next, we take a random row along the simulation, and examine the physical structures that are intercepted. Next, consider the column density distribution of the clouds. We find that the lines that are identified as absorption systems generally correspond well with overdense regions in real space. Therefore we can obtain the column density distribution

Figure 3. Thin solid and dotted lines show the column density distribution at redshift $z = (3, 2)$. Solid dots indicate the observed number of absorption lines with $N_{HI} > 10^{14} \text{cm}^{-2}$ (Carswell et al. 1991), and $N_{HI} > 10^{13.75} \text{cm}^{-2}$ and $N_{HI} > 10^{14.27} \text{cm}^{-2}$ (Rauch et al. 1992). Short dashed lines show the range of slopes of the column density distribution around $N_{HI} = 10^{14} \text{cm}^{-2}$ obtained by different observational methods (Rauch et al. 1992; Petitjean et al. 1993; Press & Rybicki 1993). Solid and dotted lines are extraplated column density distribution at $z = (3, 2)$ for $J \simeq 10^{-21}$. In addition, we show, as the thick dashed curve, the result from a $144^3$ simulation of the same model scaled to $J = 10^{-21}$, to show the numerical resolution effect.

approximately by taking, through each column of the simulation at $z = 3$ (a total of $288 \times 288 \times 3$), all the connected regions of cells with an overdensity larger than unity, considering each such region as a "cloud". The result is shown in Figure 3, at redshifts $z = (3, 2)$, as the thin solid and dotted lines respectively. Also shown as a solid square is the observed number of absorption lines with $N_{HI} > 10^{17.3} \text{cm}^{-2}$ from Sargent et al. (1989), and solid dots from Rauch et al. (1992), and an open circle from Petitjean et al. (1993). The short dashed lines show the range of slopes of the column density distribution around $N_{HI} = 10^{14} \text{cm}^{-2}$ obtained by different observational methods (Rauch et al. 1992; Petitjean et al. 1993; Press & Rybicki 1993). The column density distribution we obtain is close to but somewhat steeper than observed. The number of lines with $N > 10^{13.75} \text{cm}^{-2}$ per unit redshift is smaller than observed by a factor of $\sim 2$ (Rauch et al. 1992)

Several effects could alter our derived column density distribution. First, if the intensity of the ionizing background is lower than we have computed, the curves in Figure 3 are all shifted horizontally to higher column densities. For example, if $J = 10^{-21}$, in agreement with the observed proximity effect, then the number of lines at $N_{HI} > 10^{14} \text{cm}^{-2}$ is increased to the observed value [shown as thick solid and dotted lines for redshift z=(3,2)] and the slope of the distribution shallower at the same column density (around $N_{HI} = 10^{14} \text{cm}^{-2}$), in better accord with the observations. Second, due to the limited resolution of the simulation, the number of systems with high column densities is probably underestimated (illustrated by the thick dashed curve from a $144^3$ simulation of the same model scaled to $J = 10^{-21}$). Finally, due to the small size of our box, large-scale power is missed, and systems arising from structures on large scales are not taken into account. Correcting this should produce more systems with higher column densities, decreasing the slope of the distribution at $N_{HI} = 10^{14} \text{cm}^{-2}$.

We now examine the redshift evolution of the clouds. If the redshift evolution is parameterized as $N(z) \propto (1 + z)^\gamma$, we find, for lines with $N_{HI} \geq 10^{14} \text{cm}^{-2}$, $\gamma = (6.9, 4.1)$ over the ranges $z = (4 \to 3, 3 \to 2)$, respectively, and $\gamma = (4.9, 1.8)$ over the same ranges for $N_{HI} \geq 10^{16} \text{cm}^{-2}$. The evolution of the number of lines with redshift is sensitive to the variation of $J$. If $J$ is constant from redshift four to two, the same values of $\gamma$ change to $(3.5, 4.5)$ and $(2.5, 2.2)$, respectively. The redshift evolution appears to be steeper than observed (e.g., Lu et al. 1991), although the fact that the evolution is less pronounced at larger column densities agrees with observations. The rapid destruction of the clouds is presumably due to their expansion and to merging into larger structures. Also, different cosmological models will predict different rates of evolution.

We can also see from Figure 1, from the contours of the neutral column density, that lines with $N_{HI} \sim 10^{14} \mathrm{cm}^{-2}$ have transverse sizes of $\sim 20 h^{-1}$kpc, which agrees with the observational constraints on the physical size of the clouds from gravitationally lensed quasars (Foltz et al. 1984; Smette et al. 1992). Thus, we see that the structure found in the simulation does in fact roughly correspond to the well studied "Lyman alpha forest".

Our model can make predictions for the high signal-to-noise observations of the $Ly_\alpha$ forest with the new generation of large telescopes. We have analyzed the same 30 random spectra assuming $S/N = 60$ on a sampling bin of 3 kms$^{-1}$, and resolution of 7 kms$^{-1}$. Of 60 features detected, 46 are normal lines. Seven show broad wings which are modeled as blends of two systems at almost the same wavelength, but very different b-parameters; this is caused by high-velocity gas infalling into the system, and a distribution of temperatures (the strongest line in Fig. 2c is an example). Another 7 are "Gunn-Peterson features", with $N_{HI} < 10^{12.7} \mathrm{cm}^{-2}$ and $b > 30$ kms$^{-1}$, caused by small fluctuations in the IGM. When the signal-to-noise is reduced to 15, only two lines with broad wings could be detected, and no Gunn-Peterson feature was found, out of 37 detected lines.

Deviations from Voigt profiles are inherent to our model. When attempting to fit them with blends, they should cause a large peak in the spatial correlation of the lines at distance of order $\sim b$, which increases with signal-to-noise. A second prediction is that a fluctuating Gunn-Peterson effect is present between the lines, but this can be modeled as low column density lines.

## 5. CONCLUSIONS

We have shown that the growth of structure on subgalactic scales by gravitational collapse can probably account for the observed $Ly_\alpha$ forest. The low column density ab-sorption lines are produced by photoionized gas in regions resembling Zel'dovich pancakes, where gas is shock-heated and is maintained at an overdensity of 3 to 30 by a combination of gravity and ram pressure from infalling gas. The geometry of the low column density "clouds" varies from sheet-like to filamentary to quasi-spherical.

$Ly_\alpha$ clouds should therefore be considered as probes to the formation of structure and to the primordial density fluctuations of the universe, in addition to galaxies. The calculation of the density and temperature distribution of the neutral gas that we have done in our cosmological simulation is still subject to several uncertainties, due to the limited resolution of the simulation, the exclusion of large-scale power, and the poorly known ionization history of the intergalactic medium and the spectrum of the ionizing radiation. Some of these problems will be improved by future simulations with a higher dynamic range. Given the wealth of observational data that is available on the quasar absorption systems, we may expect that their study will provide sensitive new constraints on theories of structure formation.

This work is supported in part by NASA grant NAGW-2448 and NSF grants AST91-08103 and ASC-9318185, and the W. M. Keck Foundation. We thank Martin Rees for stimulating discussions, and Bob Carswell and John Webb for providing us with their profile fitting software. The simulation was performed on Convex-3880 in NCSA with time allocation provided by M. Norman.